\documentclass[a4paper]{article}
\title{String action with multiplet of $\Theta$-terms and the
hidden Poincare symmetry}
\author{A.A. Deriglazov\thanks{alexei@fisica.ufjf.br ~ On leave of
    absence from Dept. Math. Phys., Tomsk Polytechnical University,
    Tomsk, Russia}}
\date{Instituto de F\'\i sica, Universidade Federal do Rio de Janeiro,\\
Rio de Janeiro, Brasil.}
\begin{document}
\maketitle
\large
\begin{abstract}
We study string action with multiplet of $\Theta$-terms added, which
turns out to be closely related with the bosonic sector of $D=11$
superstring action [3,4]. Alternatively, the model can be considered 
as describing class of special solutions of the membrane. 
An appropriate set of variables is
find, in which the light-cone quantization
turns out to be possible. It is shown that anomaly terms in the
algebra of the light-cone Poincare generators are absent for the
case $D=27$.
\end{abstract}

{\bf PAC codes:} 11.25.Pm, 11.30.Pb 

\noindent
\section{Introduction}
Construction of $D=11$ Green-Schwarz type superstring action presents
a nontrivial problem already at the classical level. The reason is that
only for the dimensions $D=3,4,6,10$ the action is invariant under the
local $\kappa$-symmetry (as well as under the global supersymmetry) [1].
Recently it was recognized [2-6] that the problem can be resolved if one
introduces an additional vector variable $n^N$ into the formulation.
The corresponding $D=11$ action (which incorporates $n^N(\tau, \sigma)$
as the dynamical variable) was suggested in [3]. Similarly to the
Green-Schwarz construction, the action has $\kappa$-symmetry which allows
one to remove half of fermionic coordinates and supply free dynamics for
the physical variables as well as the discrete mass spectrum [3,4]. 
Moreover, $n^N$-independent part of spectrum
(being classified with respect to $SO(1,9)$ group) was identified with
the type IIA superstring states. For the massless level classified with 
respect to $SO(1,10)$ group one gets the supergravity multiplet in 
$D=11$ [7-9]. Other states (presented on each
mass level) may correspond to the states of the uncompactified M-theory
limit [9,10]. Due to these properties one hopes that such a kind theory
can be reasonable extension of the Green-Schwarz action to the
case $D=11$.

The aim of this work is to investigate some quantum properties of the
bosonic toy model inspired by $D=11$ superstring action (short 
version of the work is presented in [11]). The model is 
specified in Sec. 2 by mean of its own system of the Hamiltonian 
constraints in $D$-dimensional Minkowski space-time. The system contains 
all the necessary information for discussion of the light-cone quantization 
(remind also that any reparametrisation invariant free theory is  
determined unambiguously by given set of the constraints). It is 
demonstrated that the light-cone quantization is possible, which allows one 
to compute algebra of the light-cone Poincare generators. We show that 
anomaly terms in the algebra are absent for the case $D=27$. 
Note that generalisation of the present analysis to the supersymmetric case 
is straightforward since fermionic sector of the superstring action do 
not involves of extra auxiliary fields.

Lagrangian formulation for the model is discussed in Sec. 3. We present two 
different Lagrangian actions, both of them reproduce the model 
under consideration in the Hamiltonian formulation. The first action has 
only $(D-1)$-dimensional manifest Poincare invariance and represents string 
with multiplet of $\Theta$-terms added [9, 16]. The second action has 
$D$-dimensional manifest Poincare invariance and turns out to be closely 
related with the bosonic sector of $D=11$ superstring considered in [3, 4]. 

Besides the string coordinates, $D$-dimensional action involves some 
auxiliary variables (in particular, the abovementioned vector 
$n^N(\tau, \sigma)$). In Sec. 4 we discuss a possibility that these 
variables (and the corresponding terms in the action) originate from the 
membrane action \footnote{Note that role of the auxiliary variables 
in the supersymmetric version is to supply the local $\kappa$-symmetry. 
The last was established in [3, 4] by using of the same $D=11$ 
$\gamma$-matrix identity as for the supermembrane [18].}. Namely, we 
select particular class of solutions of the membrane equations of motion, 
which preserve manifest $d=2$ reparametrisation invariance. Our 
$D$-dimensional action has the same class as a general solution of 
equations of motion and thus  
can be considered as a theory which describe 
this particular sector of the membrane. Let us stress that contrary to 
the complete membrane theory [20, 21], 
the restricted version has discrete mass spectrum 
and definite critical dimension. In Conclusion we enumerate results of 
the work and discuss relation of the model considered with the bosonic 
sector of $D=11$ superstring action. 

\section{Light-cone quantization of the model and the critical dimension.} 

Besides the standard closed string coordinates 
$\tilde x^N, ~ \tilde p^N, ~ \tilde\alpha_n^N, ~ \tilde{\bar\alpha}_n^N$, 
the model involve a pair of the (real) conjugated variables 
$\tilde y^N, ~ \pi^N$, where $\pi^N$ is zero mode of the abovementioned 
vector $n^N(\tau, \sigma)$. Our starting point is   
$D$-dimensional Virasoro constraints
$(N=0,1, \ldots ,D-1)$
\begin{eqnarray}\label{1}
L_n=\frac 12\sum_{\forall k}\tilde\alpha^N_{n-k}\tilde\alpha^N_k=0, \quad 
\bar L_n=\frac 12\sum_{\forall k}\tilde{\bar\alpha}^N_{n-k}
\tilde{\bar\alpha}^N_k=0,
\end{eqnarray}
accompanied by the following second class system
\begin{eqnarray}\label{2}
\pi^N\tilde\alpha^N_n=0, \qquad \pi^N\tilde{\bar\alpha}^N_n=0, \qquad
n \ne 0;
\end{eqnarray}
\begin{eqnarray}\label{3}
\pi^N\tilde\alpha_0^N=0, \qquad \pi^N\tilde x^N=0, 
\end{eqnarray}
which implies $\pi^N\pi^N\ne 0$.
Below we will omit expressions for the left moving oscillators
$\tilde{\bar\alpha}^N$. The cases of $SO(1,D-1)$ and $SO(2,D-2)$ group
will be considered simultaneously:
$\eta^{NM}=(\eta^{\mu\nu}, \eta^{D-1,D-1}\equiv\eta), ~ \eta=\pm 1, ~ 
\eta^{\mu\nu}=(-,+,\ldots ,+), ~ \mu, \nu=0,1, \ldots ,D-2$.
The parameter $\eta$ is not fixed (except the
restrictions which follow from the constraints) throughout the work,
but is expected to be fixed in the supersymmetric version [4].
The string tension is chosen to be $T=\frac{1}{4{\bf \pi}}$ such that
$\tilde\alpha_0^N=-\tilde{\bar\alpha}_0^N=\tilde p^N$. 
Note that one more condition $\pi^2=const$ can be added to the system 
(\ref{1})-(\ref{3}) without spoiling of the subsequent analysis [11].
Spectrum is formed by action on the vacuum of oscillator modes 
only [6]. So, the sector $\tilde y^N, \pi^N$ can not produce extra 
negative norm states.
$D$-dimensional Poincare generators are realized as
\begin{eqnarray}\label{4}
{\bf P}^N=-\tilde p^N, \qquad 
{\bf J}^{MN}=
\tilde x^{\left[ M\right.} \tilde p^{\left. N\right]}
+iS^{MN}+i\bar S^{MN}+\tilde y^{\left[ M\right.}\pi^{\left. N\right]}, \cr
S^{MN}\equiv \sum_{n=1}^{\infty}\frac 1n
\tilde\alpha^{\left[ M\right.}_{-n}\tilde\alpha^{\left. N\right]}_n.
\end{eqnarray} 
Below we present and discuss two possible interpretations for the
system (\ref{1})-(\ref{3}) in the Lagrangian framework. First, 
equivalent to those of (\ref{1})-(\ref{3}) system can be reproduced 
starting from action of $(D-1)$-dimensional string with multiplet of
$D$ $\Theta$-terms added \footnote{Note that string with 
one $\Theta$-term is known to be
equivalent to $D$-string (see [16,17] and references therein), 
where it can be easily taken into
account in the path integral framework. It can be clue to
understanding of $n^N$-dependent part of spectrum.}.
While the action has only manifest $(D-1)$-dimensional
Poincare invariance, the correspondence means that it has also hidden
$D$-dimensional Poincare symmetry. We show that it can be actually
rewritten in a manifestly $D$-dimensional Poincare invariant form. 
It gives the second
interpretation: the resulting action turns out to be closely
related to the bosonic sector of $D=11$ superstring. Namely,
the system (\ref{1}),
(\ref{2}) can be obtained by means of partial fixation of gauge for
the bosonic constraints presented in the theory [6]. As it was shown
in [3,4], these constraints (and the corresponding terms in the action)
are essential for establishing of the $\kappa$-symmetry. 

Our aim now will be to perform light-cone quantization of the system
(\ref{1})-(\ref{3}). Then we show that anomaly terms in the 
light-cone Poincare algebra  
are absent for the critical dimension being $D=27$. Note that it is 
not surprising result since 
from Eq.(\ref{2}) it follows that one component of each oscillator is
nonphysical degree of freedom. So one expects that only the remaining
$D-1$ components will give contribution into the anomaly terms, such
that the condition of absence of the anomaly will be: $D-1=26$. We
support this suggestion by direct calculations.

To quantize the theory we follow to the standard prescription [12,13].
The second class constraints (\ref{2}), (\ref{3}) can be taken into
account by means of introduction of the corresponding Dirac bracket.
For our basic variables the non zero brackets turn out to be
\begin{eqnarray}\label{5}
\left\{\tilde x^N, \tilde p^M\right\}=\Pi^{NM}\equiv\eta^{NM}-
\frac{1}{\pi^2}\pi^N\pi^M,\nonumber\\
\left\{\tilde\alpha^N_n, \tilde\alpha^M_k\right\}=
in\delta_{n+k,0}\Pi^{NM},  \nonumber\\
\left\{\tilde y^N, \pi^M\right\}=\eta^{NM}, \nonumber\\
\left\{ \tilde y^N, \tilde y^M\right\}=
-\frac{1}{\pi^2}\tilde x^{\left[ N\right.}\tilde p^{\left. M\right]}-
i\sum_{n=1}^\infty\frac{1}{n\pi^2}
(\tilde\alpha^{\left[ N\right.}_{-n}\tilde\alpha^{\left. M\right]}_n+
\tilde{\bar\alpha}^{\left[ N\right.}_{-n}
\tilde{\bar\alpha}^{\left. M\right]}_n), \\
\left\{ \tilde x^M, \tilde y^N\right\}=
\frac{1}{\pi^2}\pi^M\tilde x^N, \qquad
\left\{ \tilde p^M, \tilde y^N\right\}=
\frac{1}{\pi^2}\pi^M\tilde p^N,  \nonumber\\
\left\{ \tilde\alpha^M_n, \tilde y^N\right\}=
\frac{1}{\pi^2}\pi^M\tilde\alpha^N_n\nonumber,
\end{eqnarray}
and the same expressions for the left moving oscillators
$\tilde{\bar\alpha}^N_n$. Now Eqs.(\ref{2}),(\ref{3}) can be solved
\begin{eqnarray}\label{6}
\tilde z^{D-1}=-\frac{\eta}{\pi^{D-1}}\pi^\nu \tilde z^\nu,
\end{eqnarray}
where $\tilde z=(\tilde x, \tilde p, \tilde\alpha_n,
\tilde{\bar\alpha}_n)$. Since brackets for the remaining variables
$\tilde x^\nu, \tilde p^\nu, \tilde\alpha^\nu_n, \tilde y^N, 
\pi^N$ are rather complicated,
it is convenient to simplify them by means of an appropriate variable
change. The change turns out to be 
(where $(\pi\tilde x)\equiv \pi^\nu\tilde x^\nu$ through this section)
\begin{eqnarray}\label{7}
x^\mu=\tilde x^\mu+c\pi^\mu(\pi \tilde x), \qquad
p^\mu=\tilde p^\mu+c\pi^\mu(\pi\tilde p), \cr
\alpha^\mu_n=\tilde\alpha^\mu_n+c\pi^\mu(\pi\tilde\alpha_n), \cr
y^\mu=\tilde y^\mu+c\left[(\pi\tilde x)\tilde p^\mu-
(\pi\tilde p)\tilde x^\mu\right]+ \cr
ic\sum_{n=1}^\infty\left[\frac{1}{n}
(\pi\tilde\alpha_{-n})\tilde\alpha^\mu_n+
(n\leftrightarrow -n)\right]+(\tilde{\bar\alpha}- sector), \cr
y^{D-1}\equiv\tilde y^{D-1}.
\end{eqnarray}
The factor $c$ is any solution of the equation
$\pi^2c^2+2c-\eta(\pi^{D-1})^{-2}=0$, thus
\begin{eqnarray}\label{8}
c=\frac{1}{\pi^2}\left[-1\pm\frac{(\eta\pi^N\pi^N)^
{\frac{1}{2}}}{\pi^{D-1}}\right].
\end{eqnarray}
The new variables obey to the canonical brackets
\begin{eqnarray}\label{9}
\{x^\mu, p^\nu\}=\eta^{\mu\nu}, \quad \{y^N, \pi^M\}=\eta^{NM}, \quad  
\{\alpha^\mu_n, \alpha^\nu_k\}=
in\eta^{\mu\nu}\delta_{n+k,0}.
\end{eqnarray}
Eq.(\ref{7}) is invertible, an opposite change has the same form and
can be obtained from Eq.(\ref{7}) by means of substitution
$z\leftrightarrow\tilde z, ~ y\leftrightarrow\tilde y, ~ 
c\mapsto \bar c$, where
\begin{eqnarray}\label{10}
\bar c=\frac{1}{\pi^2}\left[-1\pm{\pi^{D-1}}(\eta\pi^N\pi^N)^
{-\frac{1}{2}}\right].
\end{eqnarray}
Note that a variable change which leads to Eq.(\ref{9}) is not unique.
For example (for the Dirac bracket which corresponds to Eq.(\ref{2}))
the following simple change
\begin{eqnarray}\label{11}
\alpha^\mu_n=\tilde\alpha^\mu_n-\pi^\mu\frac{\tilde\alpha^{D-1}_n}
{\pi^{D-1}}, \qquad \alpha^\mu_{-n}\equiv\tilde\alpha^\mu_{-n}, \cr
y^N=\tilde y^N+
i\sum_{n=1}^\infty\frac{1}{n\pi^{D-1}}
(\tilde\alpha^N_{-n}\tilde\alpha^{D-1}_n+
\tilde{\bar\alpha}^N_{-n}\tilde{\bar\alpha}^{D-1}_n),
\end{eqnarray}
gives also the canonical brackets for the new variables. The problem is
that the Virasoro constraints, being rewritten in terms of these
variables, will contain products of $\alpha^-_n$ oscillators:
$L_n\sim p^+\alpha^-_n+\frac 12(\pi^+)^2\sum_{k=0}^{n-1}
\alpha^-_{n-k}\alpha^-_k+\ldots$. It do not allows one to resolve the
constraints in the light-cone gauge. In contrast, our change
(\ref{7}) leads to the "linearised" form of the constraints. Namely,
substitution of Eqs.(\ref{6}), (\ref{7}) into Eq.(\ref{1}) gives the
expressions
\begin{eqnarray}\label{12}
L_n=\frac 12\sum_{\forall k}\alpha^\mu_{n-k}\alpha^\mu_k=0, \quad 
\bar L_n=\frac 12\sum_{\forall k}\bar\alpha^\mu_{n-k}
\bar\alpha^\mu_k, \cr
L_0+\bar L_0=(p^\mu)^2+\sum_{k=1}^\infty(\alpha^\mu_{-k}\alpha^\mu_k+
\bar\alpha^\mu_{-k}\bar\alpha^\mu_k)=0,
\end{eqnarray}
\begin{eqnarray}\label{13}
L_0-\bar L_0=\sum_{k=1}^\infty(\alpha^\mu_{-k}\alpha^\mu_k-
\bar\alpha^\mu_{-k}\bar\alpha^\mu_k)=0, \qquad \mu=0,1,\ldots ,D-2
\end{eqnarray}
which contain the variables $p^\mu, \alpha^\mu_n, \bar\alpha^\mu_n$
only. Now the light-cone quantization can be carried out in the
standard form [7,14,15]. One imposes the gauge
$x^+=\alpha^+_n=\bar\alpha^+_n=0$, then the
variables $p^-, \alpha^-_n, \bar\alpha^-_n$ can be expressed through
the remaining
(D-3)-dimensional oscillators $\alpha^i_n, ~ \bar\alpha^i_n, ~ 
i=1,2,\ldots ,D-3$
\begin{eqnarray}\label{14}
p^-=\frac{1}{2p^+}(L_0^{tr}+\bar L_0^{tr}-a), \quad 
\alpha^-_n=\frac{1}{p^+}L_n^{tr}, \quad
\bar\alpha^-_n=-\frac{1}{p^+}\bar L_n^{tr}, \cr
L_n^{tr}=\frac 12\sum_{\forall k} \alpha^i_{n-k}\alpha^i_k, \qquad
L_0^{tr}=\frac 12(p^i)^2+\sum_{k=1}^{\infty}\alpha^i_{-k}\alpha^i_k.
\end{eqnarray}
The oscillators are arranged in the normal order, the corresponding
normal ordering constant $a$ is included into the expression for $p^-$.

By using of Eqs.(\ref{4}), (\ref{7}), (\ref{14}) one obtains the
light-cone Poincare generators which can be presented as
\begin{eqnarray}\label{15}
{\bf P}^\mu={\bf P}^\mu_{(D-1)}+\bar c\pi^\mu(\pi{\bf P}_{(D-1)}), 
\nonumber\\
{\bf J}^{\mu\nu}={\bf J}^{\mu\nu}_{(D-1)}+y^{\left[\mu\right.}
\pi^{\left. \nu\right]}, \nonumber\\
{\bf P}^{D-1}=\pm\eta (\eta\pi^N\pi^N)^{-\frac 12}
(\pi{\bf P}_{(D-1)}), \\
{\bf J}^{\mu D-1}=c\pi^{D-1}\pi^\nu{\bf J}^{\nu\mu}_{(D-1)}+
y^{\left[\mu\right.}\pi^{\left. D-1\right]}.\nonumber
\end{eqnarray}
The quantities ${\bf P}_ {(D-1)}, {\bf J}_{(D-1)}$ coincide with the
standard $(D-1)$-dimensional Poincare generators of the closed string
\begin{eqnarray}\label{16}
{\bf P}^\mu_{(D-1)}=-p^\mu, \qquad 
{\bf J}^{\mu\nu}_{(D-1)}=
x^{\left[\mu\right.}p^{\left. \nu\right]}
+iS^{\mu\nu}+i\bar S^{\mu\nu}, \cr
S^{\mu\nu}=\sum_{n=1}^{\infty}\frac 1n
\alpha^{\left[ \mu\right.}_{-n}\alpha^{\left. \nu\right]}_n,
\end{eqnarray} 
where it is implied that Eq.(\ref{14}) was substituted. Note that
$-M^2=({\bf P}^\mu)^2+\eta({\bf P}^{D-1})^2\equiv (p^\mu)^2$ from which
it follows that the last from Eq.(\ref{12}) actually gives
the mass formula. Thus, in terms of the new variables (\ref{7}),
$D$-dimensional Poincare generators of the theory are presented through
the usual $(D-1)$-dimensional one. It makes analysis of the anomaly
terms an easy task. By construction, commutators of the quantities
(\ref{15}) form $D$-dimensional Poincare algebra modulo to the terms
which can arise in the process of reordering of oscillators to the
normal form. The quantities (\ref{15}) have the following structure:
$A(y, \pi)+B(\pi)C_{(D-1)}(x,p,\alpha,\bar\alpha)$, where $C_{(D-1)}$
represents the generators (\ref{16}). Then structure of any commutator
is
\begin{eqnarray}\label{17}
\left[A^1(y,\pi), A^2(y,\pi)\right]+\left[A(y,\pi), B(\pi)\right]
C_{(D-1)}+ \cr 
B^1(\pi)B^2(\pi)\left[C^1_{(D-1)}, C^2_{(D-1)}\right].
\end{eqnarray}
The first two terms can not contain of ordering ambiguoutes. So the
only source of the anomaly can be commutators of $(D-1)$-dimensional
generators (\ref{16}). The dangerous commutator is known to be
$\left[ {\bf J}^{i-}_{(D-1)}, {\bf J}^{j-}_{(D-1}\right]$, which must
be zero. Its manifest form is
\begin{eqnarray}\label{18}
[{\bf J}^{i-}_{(D-1)},~{\bf J}^{j-}_{(D-1)}]=
\frac{1}{(p^+)^2}
\left[(L_0^{tr}-\bar L_0^{tr}+a)S^{ij}-
(L_0^{tr}-\bar L_0^{tr}-a)\bar S^{ij}\right.+ \cr
\left.\sum^{\infty}_{n=1}[
\frac{D-3}{12}(n-\frac{1}{n})-2n](\alpha^{\left[ i\right.}_{-n}
\alpha^{\left. j\right]}_n+
\bar\alpha^{\left[ i\right.}_{-n}\bar\alpha^{\left. j\right]}_n)\right],
\end{eqnarray}
which is actually zero on the constraint surface (\ref{13}) and under
the conditions
\begin{eqnarray}\label{19}
D=27, \qquad a=2. 
\end{eqnarray}
Note that in terms of the variables (\ref{7}) the same critical dimension 
arises
immediately in the old covariant quantization framework also, since the
no-ghost theorem can be applied without modifications to
Eqs.(\ref{12}),(\ref{13}).

\section{Lagrangian formulation for the model.}

We have established that the constraint system (\ref{1})-(\ref{3}) 
presents example of a model with the critical dimension $D=27$. So, 
it is interesting to discuss Lagrangian formulation
which reproduces this Hamiltonian system.
It is convenient to start from the partially reduced formulation
with variables (\ref{7}), since in this case there is no 
of crossing terms among the string 
coordinates and the auxiliary ones (compare (\ref{12}), (\ref{13}) 
with (\ref{1})-(\ref{3}).
Then the theory is specified by the variable set 
$x^\nu(\tau,\sigma), ~ p^\nu(\tau,\sigma), ~ y^N, ~ \pi^N$ 
and by the standard 
$(D-1)$-dimensional Virasoro constraints (\ref{12}), (\ref{13}).
It prompts to consider action of $(D-1)$-dimensional
string with multiplet of $D$ $\Theta$-terms added
\begin{eqnarray}\label{20}
S_{(D-1)}=\frac{1}{4{\bf \pi}}\int d^2\sigma 
\left[\frac{-g^{ab}}{2\sqrt{-g}}
\partial_ax^\nu \partial_bx^\nu-n^N\epsilon^{ab}\partial_aA^N_b\right],
\end{eqnarray}
where $\nu=0,1, \ldots ,D-2, ~ ~ N=0,1, \ldots ,D-1, ~ ~ 
\epsilon^{ab}=-\epsilon^{ba}, ~ \epsilon^{01}=-1$. 
All the variables obey to the periodic boundary conditions.
The Lagrangian multiplier 
$A_a^N(\tau, \sigma)$ supplies $n^N(\tau, \sigma)=\pi^N=const$ on-shell 
(alternatively, 
$U(1)^D$ gauge invariance can be used to remove all modes of
$A^N_a, ~ n^N$ except the zero one: $A^N_0=0, ~
A^N_1(\tau, \sigma)=y^N+\pi^N\tau, ~ n^N(\tau, \sigma)=\pi^N$). 
From this it follows that the action (\ref{20}) actually leads to the 
desired picture in the Hamiltonian formalism.
Being manifestly Poincare invariant in $(D-1)$ dimensions only,
the action possess hidden $D$-dimensional Poincare symmetry, 
as it is clear from Eq.(\ref{15}) (then $A_a^N, ~ n^N$ are considered 
as $D$-dimensional Lorentz vectors). So one expects that it can be 
rewritten in a manifestly $D$-dimensional Poincare invariant form. 
The relevant action is
\begin{eqnarray}\label{21}
S_D=\frac{1}{4{\bf \pi}}\int d^2\sigma \left[\frac{-g^{ab}}{2\sqrt{-g}}
D_ax^N D_bx^N-n^N\epsilon^{ab}\partial_aA^N_b\right],
\end{eqnarray}
where $D_ax^N\equiv\partial_ax^N-\xi_an^N$, and $\xi_a(\tau, \sigma)$ 
is one more auxiliary field.  
Local symmetries of the theory are 
$d=2$ reparametrizations, Weyl symmetry and the following
transformations with the parameters $\gamma, ~ \alpha^N$
\begin{eqnarray}\label{22}
\delta x^N=\gamma n^N, \qquad \delta\xi_a=\partial_a\gamma, \qquad
\delta A^N_a=\gamma\frac{\epsilon_{ab}g^{bc}}{\sqrt{-g}}D_cx^N;
\end{eqnarray}
\begin{eqnarray}\label{23}
\delta A^N_a=\partial_a\alpha^N.
\end{eqnarray}
As it should be, total number of the parameters coincide with the 
number of primary first class constraints (\ref{26}).
Let us demonstrate that the action (\ref{21}) reproduces the equations 
(\ref{1})-(\ref{3}) in the Hamiltonian formulation. 
By direct application of the Dirac algorithm one finds the
Hamiltonian
\begin{eqnarray}\label{24}
H=\int d\sigma\left\{-\frac N2\left[\frac {1}{4{\bf \pi}} p^2+4{\bf \pi} 
(\partial_1 x^N-\xi_1n^N)^2\right]\right. \cr 
\left. -N_1p^N(\partial_1 x^N-\xi_1n^N)
+\xi_0(np)+4{\bf \pi} (n\partial_1 A_0)+
+\lambda_{(g)}^{ab}p_{(g)ab}+\right. \cr
\left. +\lambda_{(\xi)a}p_{(\xi)}^a 
+\lambda_{(A)0}^N p_{(A)}^{0N}+
\lambda_{(A)1}^N(p_{(A)}^{1N}-4{\bf \pi} n^N)+
\lambda_{(n)}^N p_{(n)}^N\right\},
\end{eqnarray}
where $p_{(q)}$ is momenta conjugate to the variable $q$, and 
$\lambda_{(q)}$ are Lagrangian multipliers for the primary constraints. It 
was denoted also $N\equiv\frac {\sqrt{-g}}{g^{00}}, ~ 
N_1\equiv\frac {g^{01}}{g^{00}}$. After determining of the secondary 
constraints (there are no of tertiary constraints in the problem), the 
complete constraint system can be presented in the form 
\begin{eqnarray}\label{25}
n^N=\frac {1}{4{\bf \pi}}p_{(A)}^{1N}, \qquad
p_{(n)}^N=0, \cr
\xi_1=4{\bf \pi}\frac {(p_{(A)}^1\partial_1x)}{(p_{(A)}^1)^2}, \qquad 
\pi^1_{(\xi)}=0; 
\end{eqnarray}
\begin{eqnarray}\label{26}
p_{(g)ab}=0, \qquad p_{(\xi)}^0=0, \qquad p_{(A)}^{0N}=0;
\end{eqnarray}
\begin{eqnarray}\label{27}
\partial_1p_{(A)}^{1N}=0;
\end{eqnarray}
\begin{eqnarray}\label{28}
\left[\frac {1}{4{\bf \pi}}p^N\pm (\partial_1x^N-
\frac {(p_{(A)}^1\partial_1x)}{(p_{(A)}^1)^2}p_{(A)}^{1N})\right]^2=0, 
\quad p_{(A)}^{1N}p^N=0.
\end{eqnarray}
Note that rank of the matrix formed by the Poisson brackets of the 
constraints depends on the value of $\pi^2$. So, the sectors 
$\pi^2\ne 0$ and $\pi^2=0$ correspond to essentially different theories. 
In particular, in the case $\pi^2=0$ one finds the first class constraints 
$\pi_{\xi}^1=0, ~ (p_{(A)}^1\partial_1x)=0$ instead of the second class 
pair from Eq.(\ref{25}). We restrict our consideration to the sector 
$\pi^2\ne 0$. 
Then the constraints (\ref{25}) are of second class, while the 
remaining ones are of first class.
An appropriate gauge for Eq.(\ref{26}) is 
\begin{eqnarray}\label{29}
g^{ab}=\eta^{ab}, \quad \xi_0=0, \quad 
A^N_0=-\displaystyle\int\limits_0^\sigma d\sigma'
\left[\xi_1D_1x^N-p_{(A)}^{1N} \right].
\end{eqnarray}
Now Eqs.(\ref{25}), (\ref{26}), (\ref{29}) can be taken into account 
by means of introduction of the Dirac bracket. Then the variables  
$g^{ab}, ~ p_{(g)ab}, ~ n^N,$ ~ $p_{(n)}^N, ~ \xi_a, ~ p_{(\xi)}^a, ~  
A_0^N, ~ p_{(A)}^{0N}$ can be omitted from consideration. The Dirac 
bracket for the remaining variables $A_1^N, ~ p_{(A)}^{1N}, ~ 
x^N, ~ p^N$ coincide with the Poisson one. In the gauge chosen equations 
of motion for the sector $A_1^N, ~ p_{(A)}^{1N}$ turn out to be linear
\begin{eqnarray}\label{30}
\partial_0A_1^N=p_{(A)}^{1N}, \qquad \partial_0p_{(A)}^{1N}=0.
\end{eqnarray}
An appropriate gauge for the constraint (\ref{27}) of this sector 
is [5, 6] $\partial_1A_1^N=0$. The only remaining degrees of freedom in 
this gauge are zero modes
\begin{eqnarray}\label{31}
A^N_1(\tau,\sigma)=y^N+\pi^N\tau, \qquad 
p_{(A)}^{1N}(\tau,\sigma)=\pi^N=const.
\end{eqnarray}
Dynamics in the sector $x^N, ~ p^N$ is governed now by the equations  
\begin{eqnarray}\label{32}
\partial_0 x^N=\frac {1}{4{\bf \pi}}p^N, \qquad 
\partial_0p^N=4{\bf \pi}\Pi^N{}_M\partial_1\partial_1 x^M,
\end{eqnarray}
where $\Pi^N{}_M=\delta^N_M-\frac {1}{\pi^2}\pi^N\pi_M$. The 
remaining constraints acquire the form
\begin{eqnarray}\label{33}
\left[\frac {1}{4{\bf \pi}}p^N\pm\Pi^N{}_M\partial_1x^M\right]^2=0, 
\qquad \pi^Np^N=0.
\end{eqnarray}
In the gauge 
\begin{eqnarray}\label{34}
\pi^Nx^N=0,
\end{eqnarray}
for the last constraint, Eq.(\ref{32}) reduce to those of the usual 
string, with the well-known solution in terms of the oscillator variables. 
Being rewritten in these terms, Eqs.(\ref{33}), (\ref{34}) coincide with 
Eqs.(\ref{1})-(\ref{3}), as it was stated above.

Thus, it was established canonical equivalence of the actions 
(\ref{20}) and (\ref{21}) -they have the same physical sector. In 
particular, while the action (\ref{20}) has only manifest 
$(D-1)$-dimensional 
Poincare invariance, it possess also hidden $D$-dimensional Poincare 
symmetry, the last is given by Eqs.(\ref{15}), (\ref{16}).

\section{The model as a special sector of the membrane.}

Particular form of the action (\ref{21}) was guessed above from the 
requirement that it reproduces the desired constraint system 
(\ref{1})-(\ref{3}) in the partially fixed gauge. In this section we 
discuss a possibility to obtain this model starting from the membrane 
action. Membrane equations of motion are intrinsically non-linear, 
which do not allows one to obtain their general solution. Some special 
solutions were considered in the literature (see [18, 22] and references 
therein). In particular, there are solutions which correspond to the 
massive 
particle and to the string [18]. Semiclassical quantization of the 
membrane was 
investigated on the ground of the spherical solution [23] and of the 
toroidal one [19]. Here we select one more class which turns out to be 
useful in the present context. We look for solutions with ansatz for 
the membrane coordinate $x^N(\tau,\sigma,\rho)$ chosen in special form. 
After substitution of the ansatz into the membrane equations of motion 
they acquire the form (\ref{32}), (\ref{33}). So the action (\ref{21}) 
can be considered as describing this particular sector of the membrane 
theory.

First note that the action (\ref{21}) can be obtained from the membrane 
action [19] 
\begin{eqnarray}\label{35} 
S=\frac{1}{4{\bf \pi}}\int d^3\sigma \frac{1}{2\sqrt{-\gamma}} \left[
-\gamma^{AB}\partial_Ax^N \partial_Bx^N+1\right],
\end{eqnarray}
by means of the following formal trick. Consider $d=2$ reparametrisation 
invariant truncation of the metric ($A=(a, 2)$)
\begin{eqnarray}\label{36}
\gamma^{ab}=g^{ab}(\tau, \sigma), \quad \gamma^{a2}=-g^{ab}
\xi_b(\tau,\sigma), \quad \gamma^{22}=1+g^{ab}\xi_a\xi_b; \cr 
g^{ab}g_{bc}=\delta^a_c, \qquad \det\gamma^{AB}=\det g^{ab};
\end{eqnarray}
and of the coordinate (cylindrical membrane is considered)
\begin{eqnarray}\label{37}
x^N(\tau,\sigma,\rho)=\tilde x^N(\tau,\sigma)+
\frac {\pi^N}{\sqrt{\pi^2}}\rho,
\end{eqnarray}
where $\pi^N=const$. Substitution of Eqs.(\ref{36}), (\ref{37}) into 
Eq.(\ref{35}) gives the expression 
\begin{eqnarray}\label{38}
S=\frac{1}{4{\bf \pi}}\int d^2\sigma \frac{-g^{ab}}{2\sqrt{-g}}
\left(\partial_a\tilde x^N-\xi_a\pi^N\right)^2.
\end{eqnarray}
Further, to avoid appearance of the fixed vector $\pi^N$ in the 
formulation, one introduces the dynamical variable 
$\pi^N\longrightarrow n^N(\tau,\sigma)$. The condition 
$\partial_an^N=0$ can be incorporated into Eq.(\ref{38}) by means of the 
Lagrangian multiplier term as follows
\begin{eqnarray}\label{39}
S_D=\frac{1}{4{\bf \pi}}\int d^2\sigma \left[\frac{-g^{ab}}{2\sqrt{-g}}
(\partial_a\tilde x^N-\xi_an^N)^2-\epsilon^{ab}A_a^N\partial_bn^N\right].
\end{eqnarray}
From equations of motion $\frac {\delta S}{\delta A}=0$ one has 
$n^N(\tau,\sigma)=\pi^N=const$, as it is desired. The resulting 
action (\ref{39}) coincides with Eq.(\ref{21}).

This trick will be legitimate if the truncation (\ref{36}), (\ref{37}) 
is consistent with the membrane dynamics. This fact can be easily 
demonstrated in the Hamiltonian formulation [22, 19]. Actually, after 
partial fixation of gauge, the membrane dynamics is governed by the 
equations of motion ($\partial_A=(\partial_0, \partial_i)$)
\begin{eqnarray}\label{40}
\partial_0x^N=\frac {1}{4{\bf \pi}}p^N, \cr 
\partial_0p^N=4{\bf \pi}\partial_1\left[(\partial_2x\partial_2x)
\partial_1x^N-(\partial_1x\partial_2x)\partial_2x^N\right] \cr 
+4{\bf \pi}\partial_2\left[(\partial_1x\partial_1x)\partial_2x^N-
(\partial_1x\partial_2x)\partial_1x^N\right],
\end{eqnarray}
and by the constraints
\begin{eqnarray}\label{41}
(p\partial_ix)=0, \qquad (4{\bf \pi})^{-2}p^2+
\det(\partial_ix\partial_jx)=0.
\end{eqnarray}
Let us look for solutions with the ansatz (\ref{37}). Substitution into 
Eqs.(\ref{40}), (\ref{41}) gives the equations \footnote{It is interesting 
to note that the ansatz $x^N(\tau,\sigma,\rho)=\tilde x^N(\tau,\sigma)$ 
gives one more class of the collapsed [18] solutions. Namely, substitution 
into Eqs.(\ref{40}), (\ref{41}) leads to the tensionless string dynamics 
[24-26].}
\begin{eqnarray}\label{42}
\partial_0\tilde x^N=\frac {1}{4{\bf \pi}}p^N, \qquad 
\partial_0p^N=4{\bf \pi}\partial_1\left[\partial_1\tilde x^N-
\frac {(\pi\partial_1\tilde x)}{\pi^2}\pi^N\right], \cr 
(\pi p)=0, \quad (p\partial_1\tilde x)=0, \quad 
(4{\bf \pi})^{-2}p^2+\left(\partial_1\tilde x^N-
\frac {(\pi\partial_1\tilde x)}{\pi^2}\pi^N\right)^2=0. 
\end{eqnarray}
The last constraint implies, in particular, 
$p^N(\tau,\sigma,\rho)=\tilde p^N(\tau,\sigma)$. The resulting 
equations (\ref{42}) are equivalent to the system (\ref{32}), 
(\ref{33}), the last was obtained from the action (\ref{21}). 
General solution of the system was discussed in the previous 
section, which confirm consistency of the trick.

Note that truncation of the type (\ref{36}), (\ref{37}) can be applied 
to the supermembrane action as well. From the previous results one expects 
that the resulting supersymmetric theory has critical dimension 
$D=11$.

\section{Conclusion}

In this work we have presented example of the bosonic string-type 
model which can be quantized in the light-cone gauge and leads to the 
critical dimension $D=27$. Two canonically equivalent Lagrangian actions 
for the model were discussed, see Eq.(\ref{20}) and Eq.(\ref{21}), with 
manifest Poincare symmetry in $(D-1)$ and $D$ dimensions correspondingly.
There is analogy between the 
action (\ref{21}) and $D$-string which can be clue for understanding 
of $n^N$-dependent part of spectrum. 
It was demonstrated that $D$-dimensional action can be considered as 
a theory which describe some class of special solutions of the membrane 
equations of motion. One expects that the truncation used can be equally 
applied to the supermembrane action, which would give supersymmetric 
version for the model considered. 

Note also that analysis of spectrum in the light-cone gauge is 
expected to be more complicated as compare with the 
standard case. In the gauge considered the manifest symmetry is $SO(D-3)$ 
while the massive states should fall into representations of the little 
group $SO(D-1)$. Similar situation arise for $D=11$ membrane 
[18,19] and was analyzed in [8]. It was demonstrated that $SO(8)$ 
multiplets of the first massive level for the toroidal supermembrane 
actually fall into representations of $SO(10)$ group. We hope that the 
analogous consideration is applicable for the present case as well.

To conclude, let us comment on relation between (\ref{21}) and the
bosonic sector of $D=11$ superstring [3,4]. The only difference is
appearance of the constraint $\pi^N\partial_1x^N=0$ instead of the
last constraint from Eq.(\ref{33}), which means that Eq.(\ref{3}) is
absent. One expects that for an appropriately chosen variables all 
the previous analysis
can be repeated for this case also. The important point is that 
zero string 
modes are not restricted, so quantum states of the theory (in
particular, fields of the low-energy effective action) will be functions
of all the momentum components $p^N$, which can simplify analysis of the 
state spectrum. 
Fermionic sector of $D=11$
superstring action do not involves of extra auxiliary fields and  
consist of $D=11$ Majorana spinor only. The last can be
decomposed on a pair of the Majorana - Weyl spinors of an opposite
chirality with respect to $SO(1,9)$ group. From this fact and from
the result $D=27$ for the bosonic sector one expects that the
critical dimension for the superstring presented in [3,4] is $D=11$.

\section*{Acknowledgments.}

The work has been supported by FAPERJ and partially by
Project INTAS-96-0308 and by Project GRACENAS 
No 97-6.2-34.

\end{document}